\documentclass[a4paper,11pt]{article}
\usepackage{jheppub}

\usepackage{makecell}
\usepackage{float}

\title{\boldmath Testing AdS early dark energy with Planck, SPTpol and LSS data}

\author[a]{Jun-Qian Jiang}
\author[a,b,c,d]{Yun-Song Piao}

\affiliation[a]{School of Physics, University of Chinese Academy
of Sciences, Beijing 100049, China}

\affiliation[b]{School of Fundamental Physics and Mathematical
Sciences, Hangzhou Institute for Advanced Study, UCAS, Hangzhou
310024, China}

\affiliation[c]{International Center for Theoretical Physics
Asia-Pacific, Beijing/Hangzhou, China}

\affiliation[d]{Institute of Theoretical Physics, Chinese Academy
of Sciences, P.O. Box 2735, Beijing 100190, China}

\emailAdd{jqjiang@zju.edu.cn} \emailAdd{yspiao@ucas.ac.cn}

\abstract{The Hubble tension might be resolved by injecting a new
energy component, called Early Dark Energy (EDE), prior to
recombination. An Anti-de Sitter (AdS) phase around recombination
can make the injected energy decay faster, which thus allows a
higher EDE fraction (so larger $H_0$) while prevents degrading the
CMB fit. In this work, we test the AdS-EDE model with CMB and
Large-Scale Structure (LSS) data. Our CMB dataset consists of
low-$\ell$ part of Planck TT spectrum and SPTpol polarization and
lensing measurements, since this dataset predicts the CMB lensing
effect consistent with $\Lambda$CDM expectation. Combining it with
BAO and Pantheon data, we find the bestfit values $H_0=71.92$
km/s/Mpc and $H_0=73.29$ km/s/Mpc without and with the SH0ES
prior, respectively. Including cosmic shear and galaxy clusters
data, we have $H_0=71.87$ km/s/Mpc and $S_8=0.785$,
i.e. only $1.3\sigma$ discrepancy with direct $S_8$ measurement. }

\begin{document}
\maketitle
\flushbottom

\section{Introduction}
\label{sec:1}

Currently, the $\Lambda$CDM model is being confronted with
unsolved tensions. One is the $H_0$ tension, i.e. the discrepancy
between the direct and indirect measurements of Hubble constant
$H_0$ \cite{Verde:2019ivm,Riess:2019qba}. Based on standard
$\Lambda$CDM model, the Planck collaboration found $H_0=67.27\pm
0.60$ km/s/Mpc \cite{Planck:2018vyg}, see also recent SDSS-IV
result \cite{eBOSS:2020yzd}. In contrast, using Cepheid-calibrated
supernovae, the SH0ES found $H_0=74.03\pm 1.42$ km/s/Mpc
\cite{Riess:2019cxk}, which is $4.4\sigma$ inconsistent with
Planck's result. Different local measurements (independent of
cosmological model) have actually bring similar higher $H_0$,
which make it unlikely to be blamed on single systematic error.
Therefore, the physics beyond $\Lambda$CDM model might be
required, see
e.g.\cite{DiValentino:2020zio,DiValentino:2021izs,Perivolaropoulos:2021jda}
for recent reviews.

Another is the $S_8$ tension, i.e. the discrepancy between CMB and
Large-Scale Structure (LSS) observations, which is quantified
using $S_8 \equiv \sigma_8\sqrt{\Omega_\text{m}/0.3}$, where
$\sigma_8$ is the amplitude of matter density fluctuation at low
redshift. Based on flat $\Lambda$CDM model, the Planck
collaboration found $S_8 = 0.834 \pm 0.016$ \cite{Planck:2018vyg}.
However, the weak lensing measurements and redshift surveys prefer
lower values, e.g. $S_8=0.766^{+0.020}_{-0.014}$ for
KiDS-1000+BOSS DR12+2dfLenS \cite{Heymans:2020gsg} and
$S_8=0.737^{+0.040}_{-0.036}$ for KiDS+VIKING-450
\cite{Hildebrandt:2018yau}, see also \cite{Nunes:2021ipq}.

Recently, the Early Dark Energy (EDE) scenario
\cite{Karwal:2016vyq,Poulin:2018cxd} for resolving the $H_0$
tension has been proposed. In light of the requirement that EDE
must be non-negligible only for a short period around
matter-radiation equality and before recombination, this early
dark component can be modelled as fields with certain
phenomenological potentials, such as ultra-light axion-like
potential \cite{Poulin:2018cxd,Smith:2019ihp}, power-law potential
\cite{Agrawal:2019lmo}, also ADE \cite{Lin:2019qug}, NEDE
\cite{Niedermann:2019olb} and CEDE \cite{Karwal:2021vpk}. In
particular, it is found in Ref.\cite{Ye:2020btb}, see also
\cite{Ye:2020oix}, that the existence of an Anti-de Sitter (AdS)
phase around recombination can lift $H_0$ more effectively while
less spoil to CMB fit. In past years, the EDE models have been
extensively studied
\cite{Alexander:2019rsc,Sakstein:2019fmf,Braglia:2020bym,Lin:2020jcb,Fujita:2020ecn,Seto:2021xua,Tian:2021omz,Sabla:2021nfy,Nojiri:2021dze,Braglia:2021fxn,Allali:2021azp},
see also modified gravity models
e.g.\cite{Zumalacarregui:2020cjh,Ballesteros:2020sik,Braglia:2020auw,Odintsov:2020qzd,Dainotti:2021pqg}.

It has been noticed that the scenarios resolving the Hubble
tension will inevitably exacerbate the $S_8$ tension between CMB
and LSS. After adding LSS data, Ref.\cite{Hill:2020osr} showed
that for ultra-light axion-like EDE, the fraction of EDE is downed
to $f_\text{EDE} < 0.06$, which makes it impossible to resolve
Hubble tension, see also similar results by using Effective Field
Theory of Large-Scale Structure (EFTofLSS)
\cite{Ivanov:2020ril,DAmico:2020ods}. These seems suggest that the
EDE is being faced with the significant challenge from LSS
observations, see also \cite{Vagnozzi:2021gjh}, though it may due
to our inappropriate Bayesian analysis method
\cite{Murgia:2020ryi,Smith:2020rxx}

There are also discords inside CMB datasets. It's well-known that
there seems to be inconsistency between the high-$\ell$ and
low-$\ell$ part in Planck's TT power spectrum
\cite{Addison:2015wyg}, where the low-$\ell$ part favors higher
$H_0$ and lower $S_8$. Planck blamed it to the combined effects of
an oscillatory-like set of high-$\ell$ residuals and the deficit
in low-$\ell$ ($\ell<30$) power \cite{Planck:2016tof}.
Gravitational lensing effect can smooth sound peaks (mainly
affects high-$\ell$ part), which, however, is stronger than what
is excepted in $\Lambda$CDM model
\cite{Addison:2015wyg,Motloch:2019gux}. Generally, one quantifies
this effect by $A_L$, which equals $1$ in a self-consistent model,
but actually $A_L=1.180 \pm 0.065$ for Planck
\cite{Planck:2018vyg}. As a contrast, we have not found this over
smoothing effect in ground-base CMB observations. Actually, we see
$A_L=1.01\pm 0.11$ for ACT \cite{ACT:2020gnv}, while $A_L=0.81\pm
0.14$ SPTpol \cite{SPT:2017jdf}, which is even lower than
excepted.

As the CMB observation covering the widest sky area, Planck
measures the most precise TT spectrum in low-$\ell$ part. Ground
based CMB observations, such as SPT, ACT and POLARBEAR, focus on
the polarization power spectrum at small scale. They can cover
higher $\ell$ than Planck. Therefore,
Ref.\cite{Chudaykin:2020acu,Chudaykin:2020igl} discarded Planck's
high-$\ell$ part and chose to combine Planck low-$\ell$ TT power
spectrum and SPTpol polarization spectrum to test the EDE model.
In this way, the most debatable part of Planck data can be
avoided, and interestingly, higher $H_0$ and lower $S_8$ are
found. See also Ref.\cite{Hill:2021yec,Poulin:2021bjr} for results with ACT DR4 data.

In AdS-EDE model \cite{Ye:2020btb}, $H_0>72$ km/s/Mpc is
compatible with direct $H_0$ measurement at $1\sigma$ level.
However, it also suffers from the $S_8$ tension. Using datasets
containing full Planck data, BAO, supernova, SH0ES data, we found
$\sigma_8=0.8514$. This larger $\sigma_8$ seems inevitable due to
the positive correlation between $\Omega_mh^2$ and $H_0$
\cite{Ye:2020oix}. Thus it is significant to revisit the
constraining power of LSS on the AdS-EDE model using the different
CMB dataset that predicts the consistent CMB lensing effect
\cite{Chudaykin:2020acu,Chudaykin:2020igl}. Moreover, it is also
interesting to estimate the effect of $H_0$ prior on AdS-EDE
model. In this work, we will test AdS-EDE model with combined
Planck\_low$\ell$+SPTpol dataset, as well as LSS data.

The paper is organized as follow. We give a brief review on
AdS-EDE model in \autoref{sec:2}. The datasets and methodology are
showed in \autoref{sec:3}. Then we check the consistency of our
CMB dataset in \autoref{sec:4} and present results in
\autoref{sec:5}. We discuss our result with power spectrum in
\autoref{sec:6}. And we conclude in \autoref{sec:7}.

\section{A brief review on AdS-EDE model}
\label{sec:2}

It is well-known that the CMB observations measures the angle
$\theta^*_\text{s}$ projected on the last-scattering surface of
sound horizon (when the photon decoupled),
\begin{equation}
    \theta^*_\text{s}=\frac{r^*_\text{s}}{D^*_A},
\end{equation}
where the sound horizon
\begin{equation}
\label{eq:sound horizon} r^*_\text{s} \equiv \int_0^{t_*}
\frac{\mathrm{d}t}{a(t)} c_\text{s}(t) = \int_{z_*}^\infty
\frac{\mathrm{d}z}{H(z)}c_\text{s}(z)
\end{equation}
with $c_\text{s}$ set by the densities of baryons and matter, and
the angle distance $D^*_A$ to the last-scattering surface by
\begin{equation}
D^*_A = \int_0^{z_*}\frac{\mathrm{d}z}{H(z)} \propto
\frac{1}{H_0}.
\end{equation}
Reducing $r_\text{s}$ will lift $H_0\sim 1/r_\text{s}$, which can
be achieved by injecting certain dark component prior to
recombination \cite{Aylor:2018drw}. The corresponding dark energy
is usually called EDE, whose the parameter of state $w \approx -1$
at beginning and hereafter $w > 1/3$ around recombination to
dilute faster than radiation.

\begin{figure}[htb]
   \centerline{\includegraphics[width=0.7\linewidth]{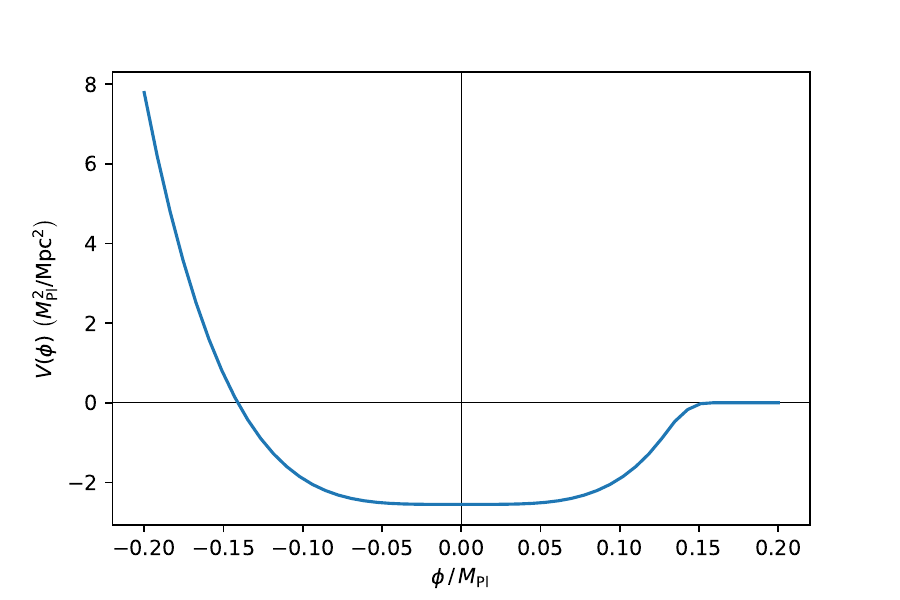}}
\caption{\label{fig: V-phi}The potential of AdS-EDE with best-fit
values to Planck\_low$\ell$ + SPT + BAO + Pantheon + SH0ES. The
field frozen at the initial point until $\partial_{\phi}^2 V(\phi)
\approx 9 H^{2}$. Then it will roll through the AdS phase, and
arrive at the flat region of potential with $V>0$ finally.}
\end{figure}

In string theory, it is easy to construct AdS vacua
e.g.\cite{Bousso:2000xa,Danielsson:2009ff}, which so are ubiquitous. And AdS vacua are
also important due to AdS/CFT duality \cite{Maldacena:1997re}.  In contrast, a
meta-stable dS vacuum is difficult to construct (i.e belongs to
the swampland) \cite{Ooguri:2006in,Obied:2018sgi,Garg:2018reu}.
See also e.g.\cite{Piao:2004me,Cai:2016thi,Li:2019ipk} 
for the implications of AdS vacua on primordial Universe
 and e.g.\cite{Dutta:2018vmq,Visinelli:2019qqu,Ruchika:2020avj,Calderon:2020hoc} 
for the implications of AdS vacua on the late Universe. 
Our phenomenological
potential modelling AdS-EDE is that in Ref.\cite{Ye:2020btb}:
\begin{equation}
    V(\phi)=\left\{\begin{array}{ll}
    V_{0}\left(\dfrac{\phi}{M_{\text{Pl}}}\right)^{4}-V_{\text{AdS}} & ,\quad \dfrac{\phi}{M_{\text{Pl}}}<\left(\dfrac{V_{\text{AdS}}}{V_{0}}\right)^{1 / 4} \\
    0 &, \quad\dfrac{\phi}{M_{\text{Pl}}}>\left(\dfrac{V_\text{AdS}}{V_{0}}\right)^{1 / 4}
    \end{array}\right.
\end{equation}
where $M_{\text{Pl}}$ is the reduced Planck mass. $V_{\text{AdS}}$
is the depth of AdS phase. See \autoref{fig: V-phi} for one of the
best-fit potential in this work.

At beginning, the field is fixed in a high-energy region of its
potential (left part in \autoref{fig: V-phi}) by Hubble fiction.
Generally, we have $\rho_{\phi}=\dot{\phi}^2 / 2+V(\phi)$,
$P_{\phi}=\dot{\phi}^2 / 2-V(\phi)$ for scalar field, so initially
$w \equiv P/\rho \approx -1$, thus the field behaves like the dark
energy. And the fraction of EDE raised with the redshift of matter
and radiation. When $H$ is deceased to satisfy $\partial_{\phi}^2
V(\phi) \approx 9 H^{2}$, the field will roll down along its
potential to AdS phase $V<0$, where $w>1$. Now $\rho_{\text{EDE}}$
will redshift as $\rho \propto a^{-3(1+w)}$. After recombination,
the field will climb up the other side of the potential and
maintains $V=const. \ge 0$, which might result in accelerated
expansion of current Universe. Here we set it to 0.

By decaying in a faster way, AdS-EDE can avoid degrading the CMB
fit, which makes higher EDE fraction (so a larger $H_0$) possible.

\section{Methodology}
\label{sec:3}

As a phenomenological model \cite{Ye:2020btb}, besides six
parameters of $\Lambda$CDM model, we choose \{$\ln(1+z_c)$,
$\omega_{\text{scf}}$, $\alpha_{\text{AdS}}$\} rather than
\{$V_0$, $V_\text{AdS}$, $\phi_\text{i}$\} as EDE's parameters.
$z_c$ is the redshift that satisfy $\partial_{\phi}^2 V(\phi_c) =
9 H_c^{2}$, i.e. the redshift when EDE field starts to slow roll,
$\omega_\text{scf}$ is the physical fraction of EDE, and
$\alpha_{\text{AdS}}$ is defined by $V_\text{AdS} =
\alpha_{\text{AdS}}\left(\rho_\text{m}(z_c)+\rho_\text{r}(z_c)\right)$.

We modified \texttt{CLASS v2.9.4} \cite{Blas:2011rf} and
\texttt{MontePython v3.4} \cite{Brinckmann:2018cvx,Audren:2012wb}
to run Markov chain Monte Carlo (MCMC) analysis, and convert EDE
parameters by shooting algorithm. Nonlinear matter power specturm
is calculated by using \texttt{HALOFIT}
\cite{astro-ph/0207664,Takahashi:2012em}, whose suitability for
EDE model is checked in \cite{Hill:2020osr}. A large
$\alpha_{\text{AdS}}$ will cause the field fails to climb up the potential
while a small $\alpha_{\text{AdS}}$ is just a run-away potential 
without AdS phase. Thus we will fix it to
a value that dose not lead to a collapsing
Universe while having a significant AdS phase. We follow \cite{Ye:2020btb} and set $\alpha_{\text{AdS}}=3.79 \times
10^{-4}$.
We also adopt the same neutrino assumption as Planck. In addition,
since the field in AdS-EDE model does not oscillate and thus
cannot be approximated using the fluid approximation, we choose to
solve the Klein-Gordon equation directly (see
\cite{Agrawal:2019lmo} for EDE with power-law potential).

Here, all parameters are assumed with flat prior. The sampling range of $\ln(1+z_c)$ is set as
$[7.5, \ 9]$ to ensure the decaying of EDE before recombination.
We also set a very small lower limit for $\omega_{\text{scf}}$ to
avoid the sampling difficulty when $\omega_{\text{scf}}$ is too
small and has too weak constraint power on other EDE
parameters.\footnote{In fact, in light of the posterior
distribution below, we need not to worry about it.} The
Gelman-Rubin criterion for all chains is converged to $R-1<0.15$.
The posterior distribution is analyzed and plotted using
\texttt{GetDist} \cite{Lewis:2019xzd}.

We consider the data sets as follows:
\begin{itemize}
    \item \textbf{Planck\_low$\ell$}: The TT power spectrum of Planck 2018 \cite{Planck:2018vyg} at $\ell<1000$, including $30 \leq \ell < 1000$ part of \texttt{Plik} likelihood and $\ell<30$ part of \texttt{Commander} likelihood. All nuisance parameters are imposed with the same prior with Planck.
    \item \textbf{SPTpol}: The polarization power spectrum (TE and EE)
measured by 500 deg$^2$ SPTpol survey \cite{SPT:2017jdf}. Their
multipole range is $50 < \ell \leq 8000$. Due to very small sky
area overlap between SPTpol and Planck, the correlation
between them can be ignored. All nuisance parameters are imposed with the same
prior with SPTpol.
    \item \textbf{SPTlen}: The CMB lensing
potential measured by 500 deg$^2$ SPTpol survey
\cite{Wu:2019hek}\cite{SPT:2019fqo}. Their multipole range is $100
< \ell \leq 2000$. Here, for simplicity, we denote the dataset
combination SPTpol + SPTlen as \textbf{SPT}.
    \item \textbf{BAO}:
low redshift BAO measured by 6DF \cite{Beutler:2011hx} in
$z=0.106$ and SDSS DR7 MGS \cite{Ross:2014qpa} in $z=0.15$, as
well as the `consensus' final result from BOSS DR12 combined
analysis \cite{BOSS:2016wmc}, their effective redshifts are
$z=0.38, 0.51, 0.61$.
    \item \textbf{Pantheon}: We take use of the magnitudes and luminosity distances data of supernovae from Pantheon. \cite{Scolnic:2017caz}
\item \textbf{SH0ES} (R19): The local $H_0$ measured with
Cepheid-calibrated supernovae by SH0ES: $H_0=74.03\pm 1.42$
km/s/Mpc \cite{Riess:2019cxk}, is regarded as a Gaussian prior.
    \item \textbf{RSD}: The `consensus' final result from BOSS DR12 \cite{BOSS:2016wmc}. RSD helps us constraint $f \sigma_8$. When we add RSD data, we also use the BAO data and the covariance of combined BAO+FS analysis.
\item \textbf{WL}: We directly take use of the constraint to $S_8$
from the combined analysis of KiDS+VIKING-450 + DES-Y1 dataset,
i.e. $S_8 = 0.755^{+0.019}_{-0.021}$ \cite{Asgari:2019fkq} for
$\Lambda$CDM. In principal, we should use the full likelihood, but
as checked in \cite{Hill:2020osr}, there is little difference in
result between a Gaussian prior on $S_8$ and full likelihood for
EDE.
\end{itemize}
In addition, we include Planck 2018 measurements of the low $\ell$
part of the EE spectrum ($\ell<30$) \cite{Planck:2018vyg} by
default in all datasets combinations. This part of data can
constrain $\tau$, so as to break the degeneracy of $A_s
e^{-2\tau}$.

\section{Consistency check of CMB data sets}
\label{sec:4}

\begin{figure}[htb]
    \centerline{\includegraphics[width=\linewidth]{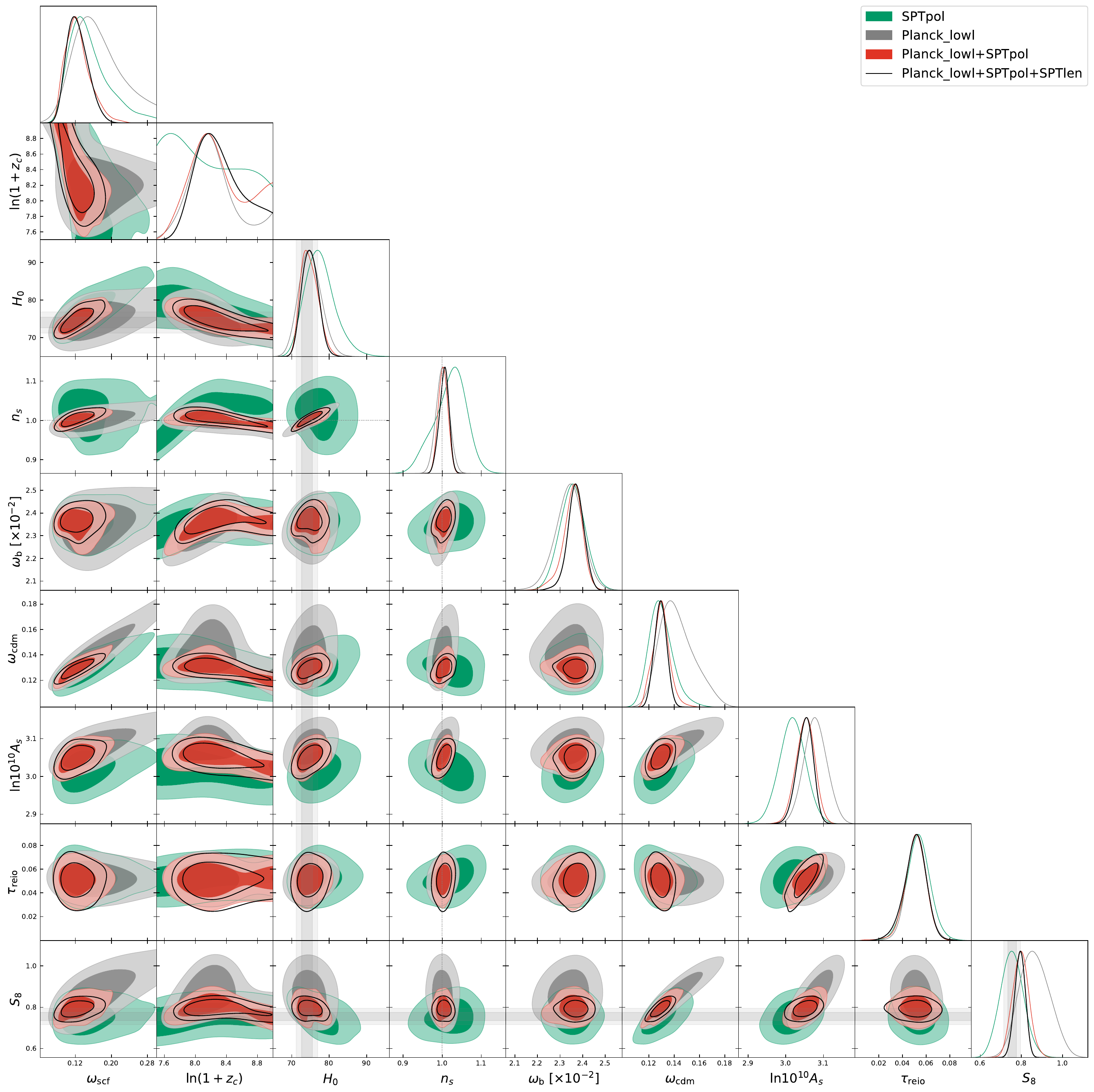}}
\caption{\label{fig:CMB consistency}Posterior distributions of CMB
data sets (68\% and 95\% confidence range). We also plot the SH0ES
constraint to $H_0$ \cite{Riess:2019cxk} and the
KiDS+VIKING-450+DES-Y1 constraint to $S_8$ \cite{Asgari:2019fkq}
in light gray.}
\end{figure}

Our CMB dataset is Planck\_low$\ell$+SPT, i.e.
Planck\_low$\ell$+SPTpol+SPTlen. Ref.\cite{Chudaykin:2020acu} has
checked the consistency of this dataset by comparing its posterior
distributions under $\Lambda$CDM model. However, as we have
modified cosmological model, we will have different posterior
distributions, some of which may be very different. Therefore, it
is required to recheck each part of CMB datasets under AdS-EDE
model. We show the corresponding posterior distributions in
\autoref{fig:CMB consistency}.

We see that they are still compatible under AdS-EDE due to large
uncertainties in individual data sets. The differences between the
cosmological parameters (including $S_8$) and AdS-EDE parameters
are within $1\sigma$ between SPTpol and Planck\_low$\ell$. The
region of the posterior distribution is reduced by the inclusion
of SPTlen, but does not change significantly. This indicates that
there is no significant conflict within the CMB data combination
Planck\_low$\ell$+SPTpol+SPTlen.

\section{Results}
\label{sec:5}

\begin{table}[]
\centering
\scriptsize
\centerline{\begin{tabular}{|c|c|c|c|c|c|}
\hline
 parameters & Planck\_low$\ell$+SPT & \makecell[c]{Planck\_low$\ell$+SPT\\+BAO+Pantheon} & \makecell[c]{Planck\_low$\ell$+SPT\\+BAO+Pantheon\\+SH0ES} & \makecell[c]{Planck\_low$\ell$+SPT\\+BAO+Pantheon\\+SH0ES+RSD+WL} \\
\hline
 $\displaystyle \omega _{\text{scf}}$ & $\displaystyle 0.125( 0.119)_{-0.027}^{+0.020}$ & $\displaystyle 0.111( 0.099)_{-0.021}^{+0.010}$ & $\displaystyle 0.122( 0.116)_{-0.021}^{+0.014}$ & $\displaystyle 0.1040( 0.1016)_{-0.014}^{+0.0088}$ \\
$\displaystyle \ln( 1+z_{c})$ & $\displaystyle 8.30( 8.25)_{-0.37}^{+0.22}$ & $\displaystyle 8.52( 8.47) \pm 0.24$ & $\displaystyle 8.39( 8.30)_{-0.23}^{+0.17}$ & $\displaystyle 8.69( 8.78)_{-0.12}^{+0.27}$ \\
$\displaystyle H_{0}$ & $\displaystyle 74.9( 74.7) \pm 2.1$ & $\displaystyle 72.24( 71.92)_{-1.3}^{+0.86}$ & $\displaystyle 73.08( 73.29) \pm 0.96$ & $\displaystyle 72.35( 71.87)_{-0.90}^{+0.69}$ \\
$\displaystyle n_{s}$ & $\displaystyle 1.005( 1.007)_{-0.011}^{+0.013}$ & $\displaystyle 0.9908( 0.9936) \pm 0.0092$ & $\displaystyle 0.9965( 0.9990)_{-0.0065}^{+0.0083}$ & $\displaystyle 0.9893( 0.9875)_{-0.0087}^{+0.0076}$ \\
$\displaystyle \omega _{\text{b}}$ & $\displaystyle 0.02369( 0.02392) \pm 0.00036$ & $\displaystyle 0.02358( 0.02386) \pm 0.00028$ & $\displaystyle 0.02368( 0.02375)_{-0.00027}^{+0.00033}$ & $\displaystyle 0.02368( 0.02367) \pm 0.00026$ \\
$\displaystyle \omega _{\text{cdm}}$ & $\displaystyle 0.1292( 0.1292) \pm 0.0048$ & $\displaystyle 0.1292( 0.1285)_{-0.0046}^{+0.0035}$ & $\displaystyle 0.1316( 0.1308) \pm 0.0041$ & $\displaystyle 0.1253( 0.1246)_{-0.0027}^{+0.0021}$ \\
$\displaystyle \ln 10^{10} A_{s}$ & $\displaystyle 3.051( 3.059)_{-0.019}^{+0.023}$ & $\displaystyle 3.042( 3.042)_{-0.018}^{+0.021}$ & $\displaystyle 3.045( 3.046)_{-0.016}^{+0.021}$ & $\displaystyle 3.025( 3.026)_{-0.015}^{+0.020}$ \\
$\displaystyle \tau $ & $\displaystyle 0.0507( 0.0545)_{-0.0082}^{+0.010}$ & $\displaystyle 0.0480( 0.0489)_{-0.0073}^{+0.0094}$ & $\displaystyle 0.0462( 0.0470)_{-0.0072}^{+0.011}$ & $\displaystyle 0.0457( 0.0456)_{-0.0074}^{+0.010}$ \\
\hline
 $\displaystyle 100\theta _{s}$ & $\displaystyle 1.04013( 1.04052) \pm 0.00083$ & $\displaystyle 1.03946( 1.03991) \pm 0.00070$ & $\displaystyle 1.03955( 1.03965) \pm 0.00073$ & $\displaystyle 1.03932( 1.03890) \pm 0.00070$ \\
$\displaystyle \sigma _{8}$ & $\displaystyle 0.829( 0.834)_{-0.015}^{+0.023}$ & $\displaystyle 0.822( 0.823) \pm 0.018$ & $\displaystyle 0.831( 0.832)_{-0.013}^{+0.017}$ & $\displaystyle 0.803( 0.801)_{-0.012}^{+0.010}$ \\
$\displaystyle S_{8}$ & $\displaystyle 0.793( 0.799) \pm 0.025$ & $\displaystyle 0.814( 0.818) \pm 0.020$ & $\displaystyle 0.820( 0.816)_{-0.017}^{+0.020}$ & $\displaystyle 0.784( 0.785) \pm 0.013$ \\
 \hline
\end{tabular}}
\caption{\label{tab:data limit}Constraints on parameters under
AdS-EDE for each data set, including mean values and $\pm 1\sigma$
regions, with best-fit values in parentheses.}
\end{table}

\subsection{Planck\_low$\ell$ + SPT}

The constraints of our combinated Planck\_low$\ell$+SPT dataset on
the parameters are shown in the thick black solid line in
\autoref{fig:CMB consistency} and \autoref{tab:data limit}. A
distinct non-Gaussian distribution can be found in the
$\ln\left(1+z_c\right)$ - $\omega_{\text{scf}}$ plane, which is
because in the low $z_c$ and low $\omega_{\text{scf}}$ region
$\theta_{\text{i}}$ is so small that the field would not climb out
of AdS well (we called it AdS bound). However, our best-fit point
is not close to this region, so we do not need to worry it. $H_0$
and $\omega_\text{scf}$ are positively correlated, which is
exactly what we expect, i.e. larger EDE fraction brings a larger
$H_0$. Meanwhile, $\omega_\text{scf}$ and $S_8$ are negatively
correlated, since larger $\omega_\text{cdm}$ is required to
balance the extra EDE, we'll discuss it in detail in
\autoref{sec:6}. The negative correlation between $z_c$ and $H_0$
is due to very little effect of the EDE peak on $r_s$ when EDE is
far away from the recombination time.

Unlike other EDE solutions (such as \cite {Chudaykin:2020igl} for
the same dataset), CMB data alone has strong preference of
non-zero EDE, at least for our model and dataset
combination.\footnote{But note that this is actually a reflection
of AdS bound. Therefore we use best-fit values as our
main result and mean values as assistance.} The posterior
distributions of $H_0$ and $S_8$ are all consistent with direct
measurements at $1\sigma$ level. The best value of $H_0$ is even
slightly larger than R19. However, the uncertainty of $H_0$ is
very large.

\subsection{Planck\_low$\ell$ + SPT + BAO + Pantheon}

\begin{figure}[ht]
    \centerline{\includegraphics[width=\linewidth]{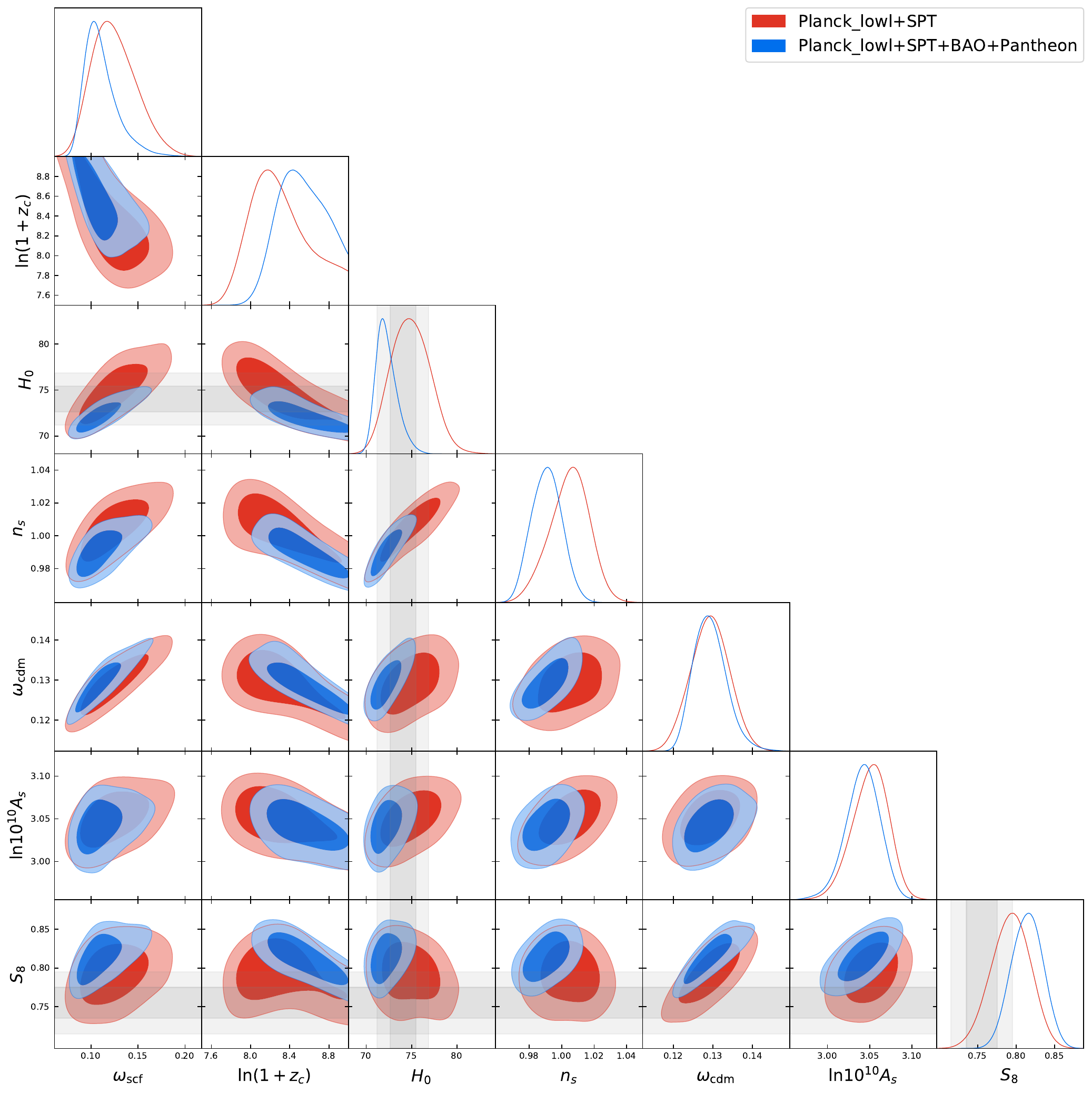}}
\caption{\label{fig:AdS-PlSBP}Planck\_low$\ell$+SPT
(+BAO+Pantheon) constraints on each parameter under AdS-EDE (68\%
and 95\% confidence range). We also plot the R19 constraint to
$H_0$ \cite{Riess:2019cxk} and the KiDS+VIKING-450+DES-Y1
constraint to $S_8$ \cite{Asgari:2019fkq} in light gray.}
\end{figure}

BAO measured $H(z)r_s^{\text{drag}}$ (line of sight direction) and
$D_M(z)/r_s^{\text{drag}}$ (perpendicular to the line of sight
direction), or average angle, constrained to
$D_V(z)/r_s^{\text{drag}}$ for the late Universe, and
(uncalibrated) supernova luminosity distances constrain the shape
of $D_L(z)$, and thus the shape of $H(z)$. Their combination sets
the constraint for $H_0 r_s^{\text{drag}}$. Recent studies have
already pointed out that the Hubble tension is actually relevant
with the deviations of both $H_0$ and $r_s$,
e.g.\cite{Knox:2019rjx,Lyu:2020lwm,Haridasu:2020pms}, so it is
necessary to compare the compatibility of model with BAO+SN
observations. We add the data of BAO+Pantheon in our
Planck\_low$\ell$+SPT dataset and present the posterior
distribution of the parameters in \autoref{fig:AdS-PlSBP}, see
also \autoref{tab:data limit}.

With the addition of BAO+Pantheon, the uncertainty of the
parameters is much reduced, however the difference from the
results without BAO+Pantheon stays within $1\sigma$. This is an
indication that BAO+Pantheon is compatible with our CMB dataset.
$H_0=71.92$ is consistent with R19 at $1.1\sigma$ level, and is
$3.7\sigma$ higher than that in $\Lambda$CDM model. The value of
$\sigma_8$ is slightly raised, but is still $2.1\sigma$ consistent
with the KiDS+VIKING-450+DES-Y1 (WL) result. The best-fit values
of EDE fraction is slightly decreased after the addition of
BAO+Pantheon.

%however, its non-zero evidence increased from $4.6\sigma$ to
%$5.3\sigma$ due to the reduction of uncertainty.

\begin{figure}[htb]
    \centerline{\includegraphics[width=1.1\linewidth]{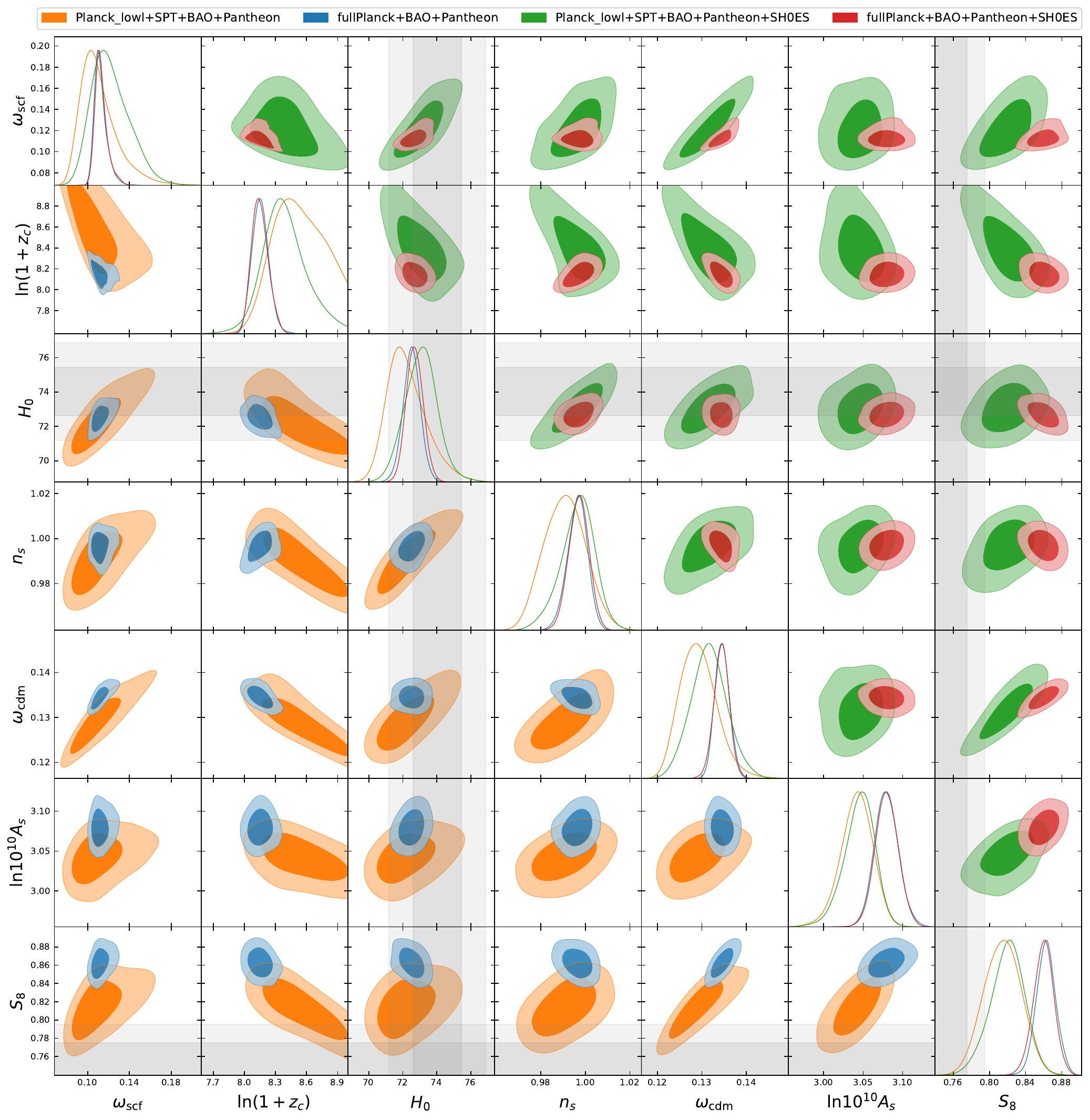}}
\caption{\label{fig:CMB compare_2d}Constraints on each parameter for
different combinations of CMB datasets + BAO + Pantheon (+SH0ES)
under AdS-EDE (68\% and 95\% confidence range). We also plot the
R19 constraint to $H_0$ \cite{Riess:2019cxk} and the
KiDS+VIKING-450+DES-Y1 constraint to $S_8$ \cite{Asgari:2019fkq}
in light gray.}
\end{figure}

We also compare the posterior distribution of the parameters in
\autoref{fig:CMB compare_2d} for our CMB dataset with those using the full
Planck data (i.e., Planck2018 high-$\ell$ TT TE EE spectrum,
low-$\ell$ TT EE spectrum and lensing spectrum) as CMB data, see
also \autoref{tab:fullPlanck}. The full Planck data has less
parameter uncertainties, but its $S_8$ differs from the WL
measurements by $4.9\sigma$. $H_0$ in the full Planck data is
slightly larger than that in our dataset, however, our EDE
parameters, namely $\ln(1+z_c)$, $\omega_{\text{scf}}$, have
larger uncertainties.

\begin{figure}[htb]
    \centerline{\includegraphics[width=\linewidth]{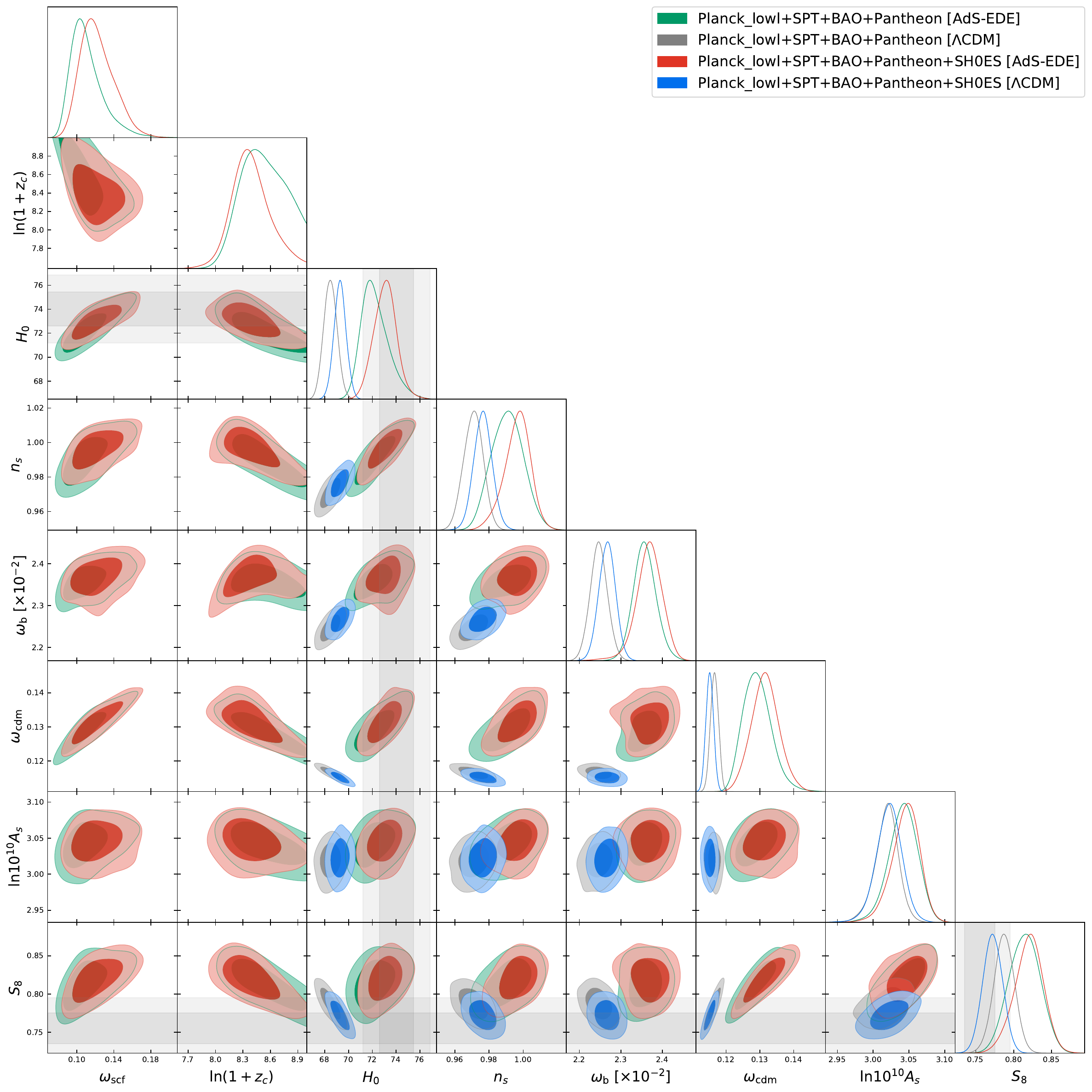}}
\caption{\label{fig:AdS-LCDM}Constraints on the parameters by
Planck\_low$\ell$+SPT+BAO+Pantheon (+SH0ES) under standard
$\Lambda$CDM and AdS-EDE (68\% and 95\% confidence range). We also
plot the R19 constraint to $H_0$ \cite{Riess:2019cxk} and the
KiDS+VIKING-450 + DES-Y1 constraint to $S_8$ \cite{Asgari:2019fkq}
in light gray.}
\end{figure}

\begin{figure}[htb]
    \centering
   \includegraphics[width=.7\linewidth]{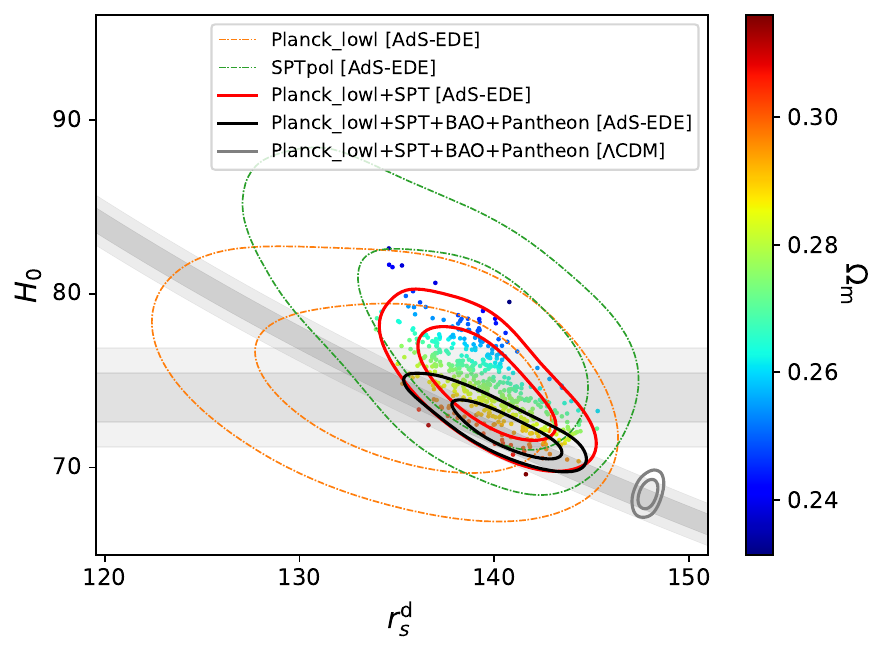}
\caption{\label{fig:rs-H0-m}$1\sigma$ and $2\sigma$ posterior
distributions for different model and dataset combinations in the
$r_s^\text{drag}$-$H_0$ plane. The horizontal gray line is the R19
constraint on $H_0$ \cite{Riess:2019cxk}. The skewed gray line is
the BAO+Pantheon constraint on $H_0 r_s^{\text{drag}}$ in a
model-independent way. The middle colored scatters are the
parameters ranges in AdS-EDE constrained by
Planck\_low$\ell$+SPT+BAO+Pantheon.}
\end{figure}

To further clarify the role of BAO+Pantheon, we follow
\cite{Bernal:2016gxb}\cite{Aylor:2018drw} by approximating $H(z)$
as a five-point natural cubic spline function and fitting the
BAO+Pantheon data with MCMC, constrain $\beta_{\text{ BAO}} \equiv
c/\left(H(z)r_s^{\text{darg}}\right)$: $\beta_{\text{BAO}}
=29.769_{-0.372}^{+0.379}$. This constraint is model-independent
(except for the flatness assumption). We compare the parameter
constraints on the $r_s^\text{darg}$-$H_0$ plane under AdS-EDE
with model-independent BAO+Pantheon in \autoref{fig:rs-H0-m}.

It can be seen that the $r_s^{\text{drag}}-H_0$ distribution under
AdS-EDE (without BAO+Pantheon data) is $1\sigma$ consistent with
the overlap region of model-independent $r_s^\text{drag}$-$H_0$
constraint and R19. However, for $\Lambda$CDM (see e.g.
\cite{Knox:2019rjx}), its posterior distribution is far from the
overlap region, and the degeneracy direction is orthogonal to the
distribution of the model-independent constraint. Similar to
$\Lambda$CDM, the direction of $\Omega_\text{m}$ in AdS-EDE is
orthogonal to the model-independent $r_s^\text{drag}$-$H_0$
constrained degeneracy direction. This can be understood as
following. It is well-known that CMB almost fixed $\theta_s$, and
\begin{equation}
    \theta_s = \frac{r_s}{D_A} = \frac{r_s}{\int \frac{\mathrm{d}z}{H(z)}} = \frac{r_s H_0}{\int
    \frac{\mathrm{d}z}{\sqrt{\Omega_\text{m}(1+z)^3+(1-\Omega_\text{m})}}}.
\end{equation}
This suggests that $\Omega_\text{m}$ will be determined by $r_s
H_0$. Therefore, the addition of BAO+Pantheon data constrains
$\Omega_\text{m}$ to larger values, so $S_8$.

\begin{table}
\centering
\begin{tabular}{|c|c|c|}
\hline
 Data set & $\Lambda$CDM & AdS-EDE \\
\hline
\hline
 Planck TT, $\displaystyle 30\leqslant \ell < 1000$ & 405.96 & 409.71 \\
Planck TT, $\displaystyle \ell < 30$ & 22.35 & 20.38 \\
Planck EE $\displaystyle \ell < 30$ & 395.68 & 395.92 \\
SPTpol & 145.56 & 144.92 \\
SPTlen & 4.91 & 4.97 \\
BOSS DR12 BAO & 3.55 & 3.60 \\
BAO low-$\displaystyle z$ & 2.30 & 2.09 \\
Pantheon & 1026.91 & 1027.27 \\
\hline
 Total & 2007.23 & 2008.86 \\
 \hline
\end{tabular}
\caption{\label{tab:chi2 PlSBP}$\chi^2$ of $\Lambda$CDM and
AdS-EDE for the Planck\_low$\ell$+SPT+BAO+Pantheon best-fit
values.}
\end{table}

We present in \autoref{tab:chi2 PlSBP} the $\chi^2$ of the best
fits of $\Lambda$CDM and AdS-EDE. The AdS-EDE fit is slightly
worse than the fit in $\Lambda$CDM: $\Delta \chi^2 = 1.63$.
However, since the difference in $\chi^2$ is slight, it is not
statistically significant.

\subsection{Planck\_low$\ell$ + SPT + BAO + Pantheon + SH0ES}

Including the R19 prior, we present the posterior parameters in
\autoref{tab:data limit}. We also present a comparison of the
parameter posterior distribution before and after the addition of
R19 data and with $\Lambda$CDM model in \autoref{fig:AdS-LCDM},
and a comparison with the parameter posterior distribution using
the full Planck data as CMB data instead in \autoref{fig:CMB compare_2d}.

We see that $H_0$ is restricted to a higher value of $H_0=73.29$
(consistent with R19 within $1\sigma$), and $S_8=0.816$. However,
the uncertainty of $H_0$ does not enlarge significantly, which
indicates that there is no significant inconsistency with R19
data. Thanks to the larger uncertainty of $H_0$ toward higher
values without R19 prior for Planck\_low$\ell$+SPT dataset, $H_0$
is raised more than that for full Planck data (see Appendix
\autoref{sec:B}). Here, $\omega_\text{scf}$ is slightly raised and
$z_c$ is constrained to be closer to the recombination epoch to
lift $H_0$. However, this constraint on the EDE parameters is not
as drastic as other EDE models, since AdS-EDE already brings a
larger $H_0$ even without the $H_0$ prior.

\begin{table}
\centering
\begin{tabular}{|c|c|c|}
\hline
 Data set & $\Lambda$CDM & AdS-EDE \\
\hline
\hline
 Planck TT, $\displaystyle 30\leqslant \ell < 1000$ & 407.06 & 407.97 \\
Planck TT, $\displaystyle \ell < 30$ & 20.95 & 20.34 \\
Planck EE $\displaystyle \ell < 30$ & 395.70 & 396.27 \\
SPTpol & 142.93 & 145.07 \\
SPTlen & 5.65 & 5.09 \\
BOSS DR12 BAO & 7.96 & 4.31 \\
BAO low-$\displaystyle z$ & 4.39 & 2.89 \\
Pantheon & 1027.46 & 1027.04 \\
SH0ES & 8.96 & 0.27 \\
\hline
 Total & 2021.07 & 2009.25 \\
 \hline
\end{tabular}
\caption{\label{tab:chi2 PlSBPS}$\chi^2$ of $\Lambda$CDM and
AdS-EDE best fits to Planck\_low$\ell$+SPT+BAO+Pantheon+SH0ES.}
\end{table}

Similarly, we present the $\chi^2$ of the $\Lambda$CDM and AdS-EDE
best-fit values with R19 data in \autoref{tab:chi2 PlSBPS}. After
the addition of R19 data, the fit of $\Lambda$CDM to the BAO and
Pantheon is bad due to the previously described deviation in the
$r_s^\text{darg}$-$H_0$ plane. However, AdS-EDE is fully
compatible with the $H_0$ prior. And due to its reduced $r_s$, the
fit to the BAO and Pantheon measurements is also well. Due to the
rapid decay of AdS-EDE, the spoil to the CMB fit is slight.
Totally, we get $\Delta \chi^2 = -11.82$ for AdS-EDE compared to
$\Lambda$CDM, indicating that AdS-EDE significantly fits relevant
observations better than $\Lambda$CDM.

\subsection{Planck\_low$\ell$ + SPT + BAO + Pantheon + SH0ES + RSD + WL}

\begin{figure}[ht]
   \centerline{\includegraphics[width=\linewidth]{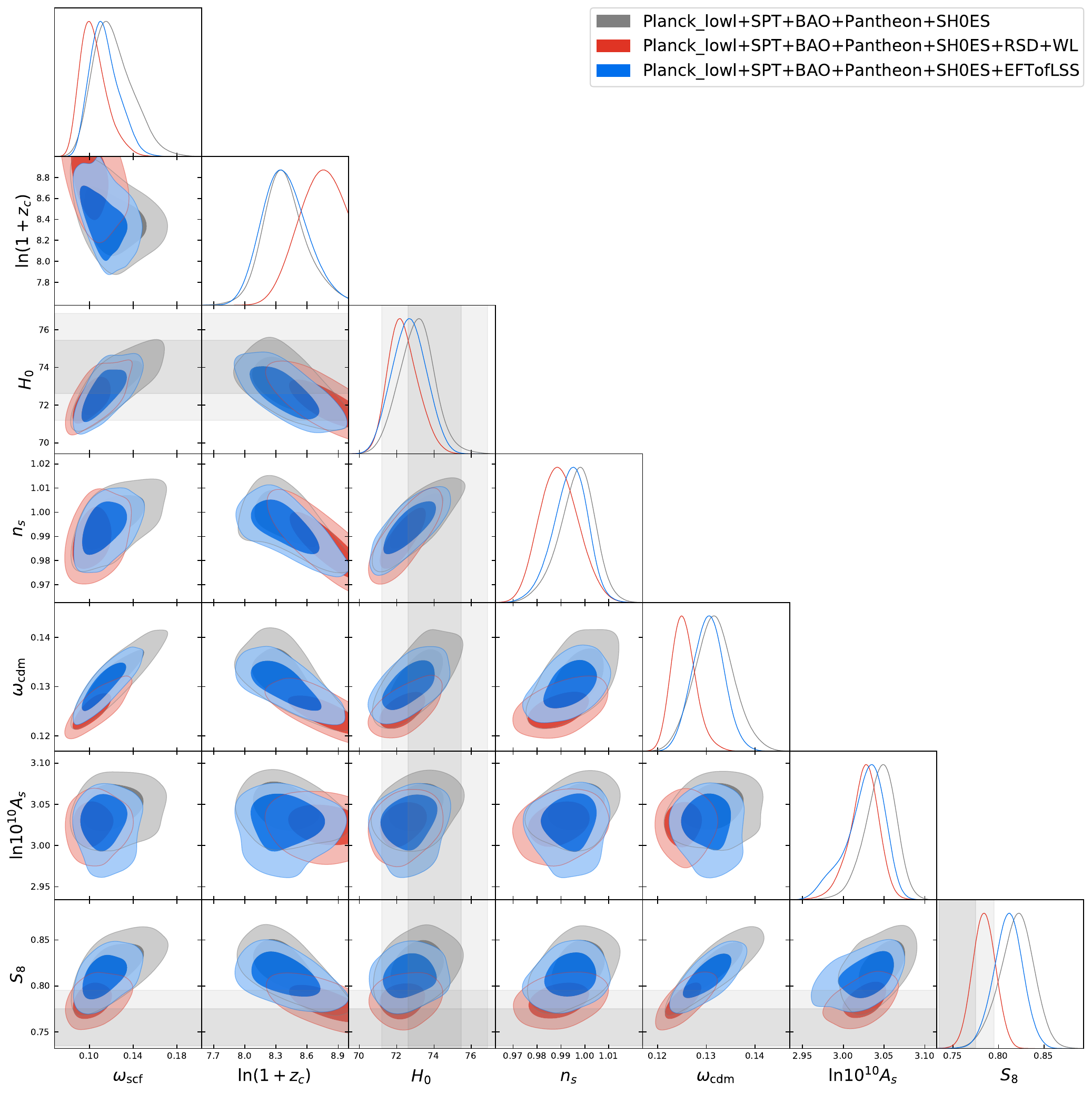}}
\caption{\label{fig:AdS_LSS}Posterior distribution of each
parameter under AdS-EDE after adding RSD and WL measurements (68\%
and 95\% confidence range). We also plot the R19 constraint to
$H_0$ \cite{Riess:2019cxk} and the KiDS+VIKING-450+DES-Y1
constraint to $S_8$ \cite{Asgari:2019fkq} in light gray.}
\end{figure}

Finally, we add RSD and WL measurements to make the relationship
with direct $S_8$ measurements clear. Results are presented in
\autoref{fig:AdS_LSS} and \autoref{tab:data limit}.

Including RSD+WL, we get $S_8=0.785$, the difference with direct
$S_8$ measurement is only $1.3\sigma$, while we have $H_0=71.87$,
which is only slightly lower compared to that without RSD+WL.
Compared with the result of full Planck data,
Planck\_low$\ell$+SPT allows higher $z_c$, so that less
$\omega_\text{cdm}$ is added to balance EDE, allowing to achieve a
lower $S_8$. This is different from compromise of lowering EDE
fraction when the full Planck data is considered.
%Our
%$\omega_\text{scf}$ still holds at $0.1016$ with $7\sigma$ for
%non-zero EDE, i.e.,
Our results indicate that AdS-EDE is not excluded by the current
WL measurements. As a contrast, we also present the result using
EFTofLSS in Appendix \autoref{sec:A}.

\section{Discussion}
\label{sec:6}

\begin{figure}[tbp]
\centering
\includegraphics[width=.32\textwidth]{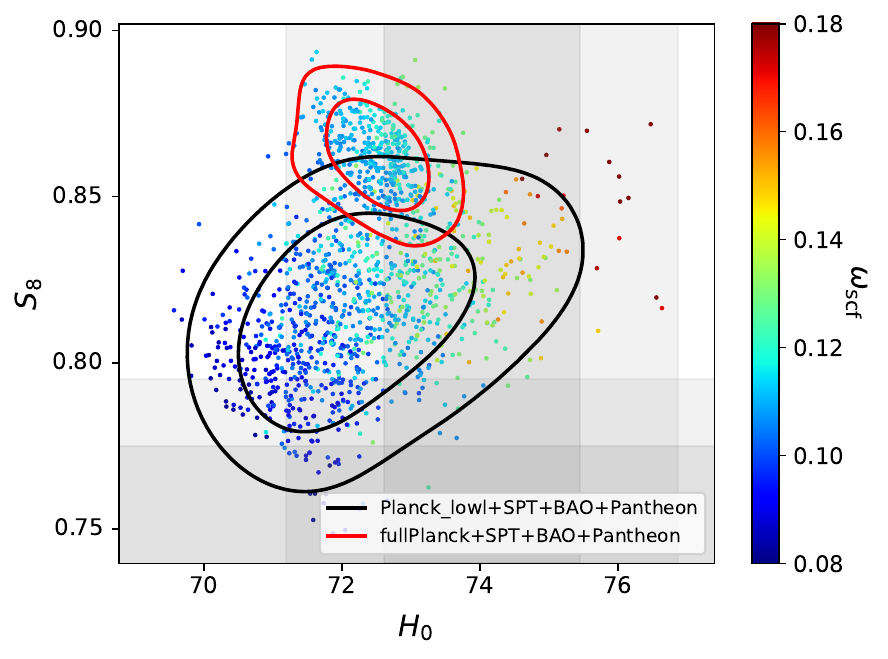}
\hfill
\includegraphics[width=.32\textwidth]{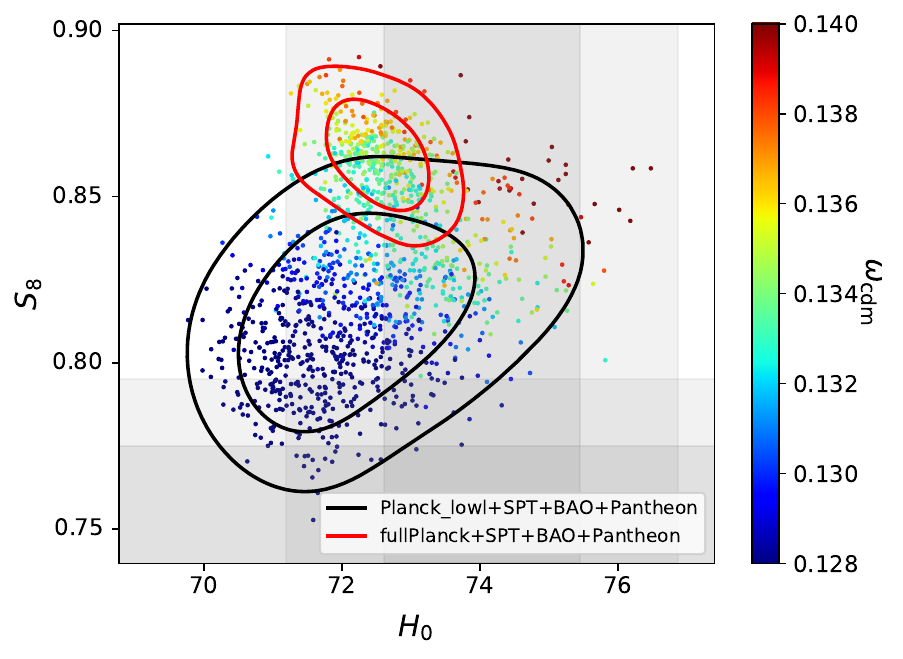}
\hfill
\includegraphics[width=.32\textwidth]{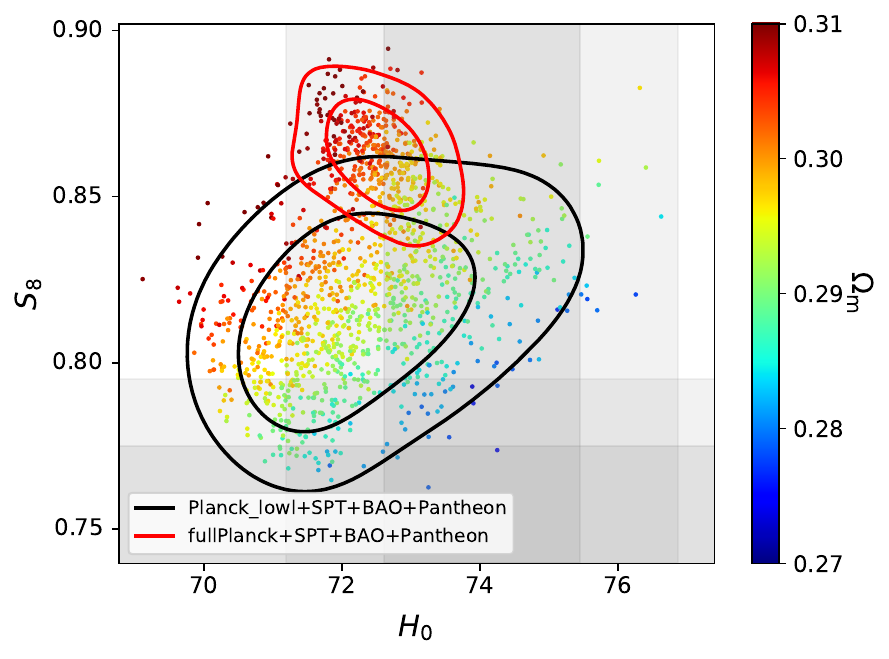}
\caption{\label{fig:H0-S8} Parameters distribution on the plane
$(S_8, H_0)$ for Planck\_low$l$+SPT+BAO+Pantheon and
fullPlanck+SPT+BAO+Pantheon in AdS-EDE. The gray bands represents
constraints from direct $H_0$ \cite{Riess:2019cxk} and $S_8$
\cite{Asgari:2019fkq} measurements.}
\end{figure}

The $S_8$-$H_0$ relation for our datasets is presented in
\autoref{fig:H0-S8}. We can find that EDE has a different
relationship from $\Lambda$CDM (see e.g.\autoref{fig:AdS-LCDM}).
Precise CMB $\theta_s^*$ measurement constrains $\omega_\text{cdm}
H_0^{1.4}$ and the heights of peaks constrain $\omega_m$, and they
actually constrain $\omega_\text{cdm} H_0$ \cite{Planck:2018vyg}. $\omega_\text{cdm}$
controls the late matter fluctuation, this is the reason of the
negative relation in \autoref{fig:AdS-LCDM}.

However, this is not the case for EDE, since the relation of
$S_8$-$H_0$ is mainly controlled by EDE's fraction
$\omega_\text{scf}$, see left plane of \autoref{fig:H0-S8}. In EDE
scenario, since larger EDE fraction brings higher $H_0$, larger
$\omega_\text{cdm}$ required to balance early ISW effect (see
middle panel of \autoref{fig:H0-S8}), which is common in any early
resolution (relevant with energy injection) for the Hubble
tension. And the approximate relationship is $\frac{\delta
H_{0}}{H_{0}} \simeq 0.5 \frac{\delta
\omega_\text{cdm}}{\omega_\text{cdm}}$ \cite{Ye:2021nej}. It is
this higher $\omega_\text{cdm}$ that bring larger $S_8$.

\begin{figure}[htb]
    \centering
   \includegraphics[width=\linewidth]{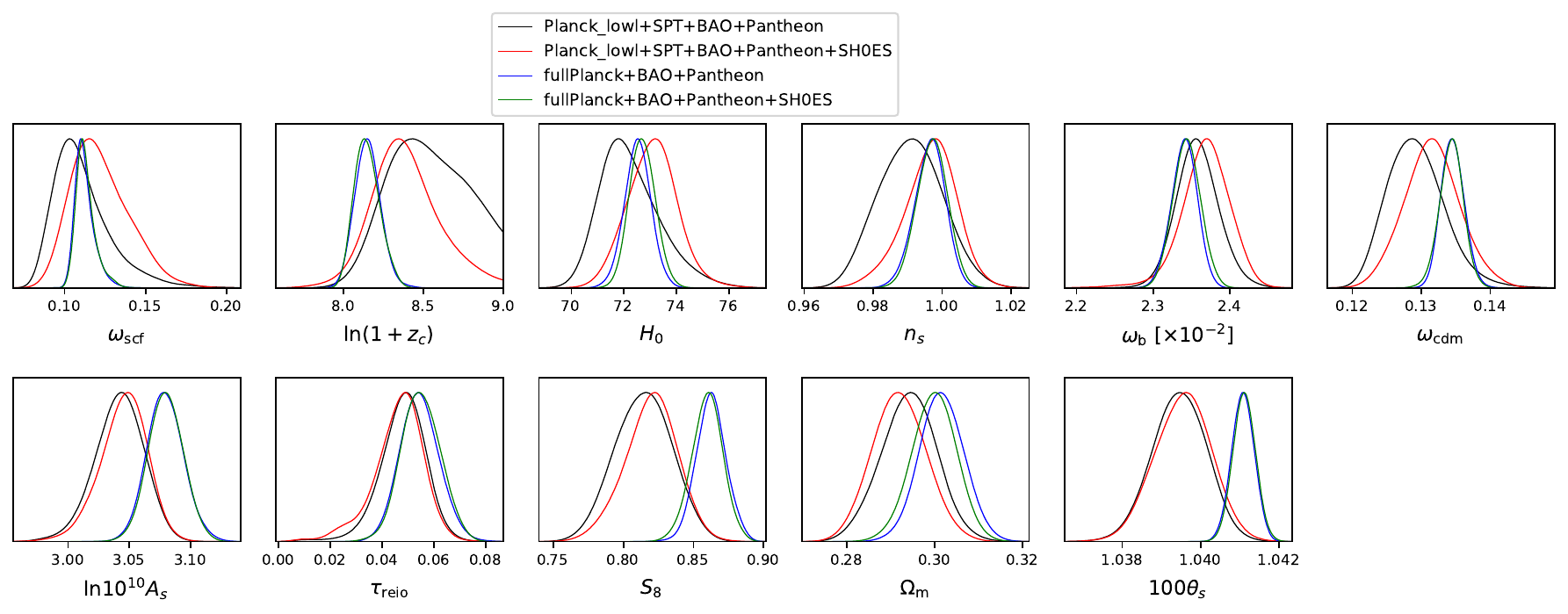}
\caption{\label{fig:CMB compare_1d}Posterior distribution of
parameters in AdS-EDE using different CMB datasets.}
\end{figure}

Besides difference with $\Lambda$CDM, there is also difference
between different CMB datasets, where Planck\_low$\ell$+SPT favors
the region consistent with direct $H_0$ and $S_8$ measurements.
This is relevant with $\Omega_m=\omega_\text{m} h^{-2}$ (see the
right panel of \autoref{fig:H0-S8}). As we have mentioned, BAO+SN
data will precisely set $\Omega_m$. However, this is the case only
when $\theta_s$ is fixed. It can be found in
\autoref{fig:CMB compare_1d} that Planck\_low$\ell$+SPT favors a
smaller $\theta_s$, which will bring less $\Omega_m$. It seems
that the slightly less $\Omega_m$ than Planck has been indicated
\cite{Moresco:2016mzx,Vagnozzi:2017ovm,DES:2017myr,Zubeldia:2019brr,KiDS:2020suj}.
Here, the change of $n_s$ is the natural results of higher $H_0$
\cite{Ye:2021nej}.

\begin{figure}[htb]
    \centering
   \includegraphics[width=\linewidth]{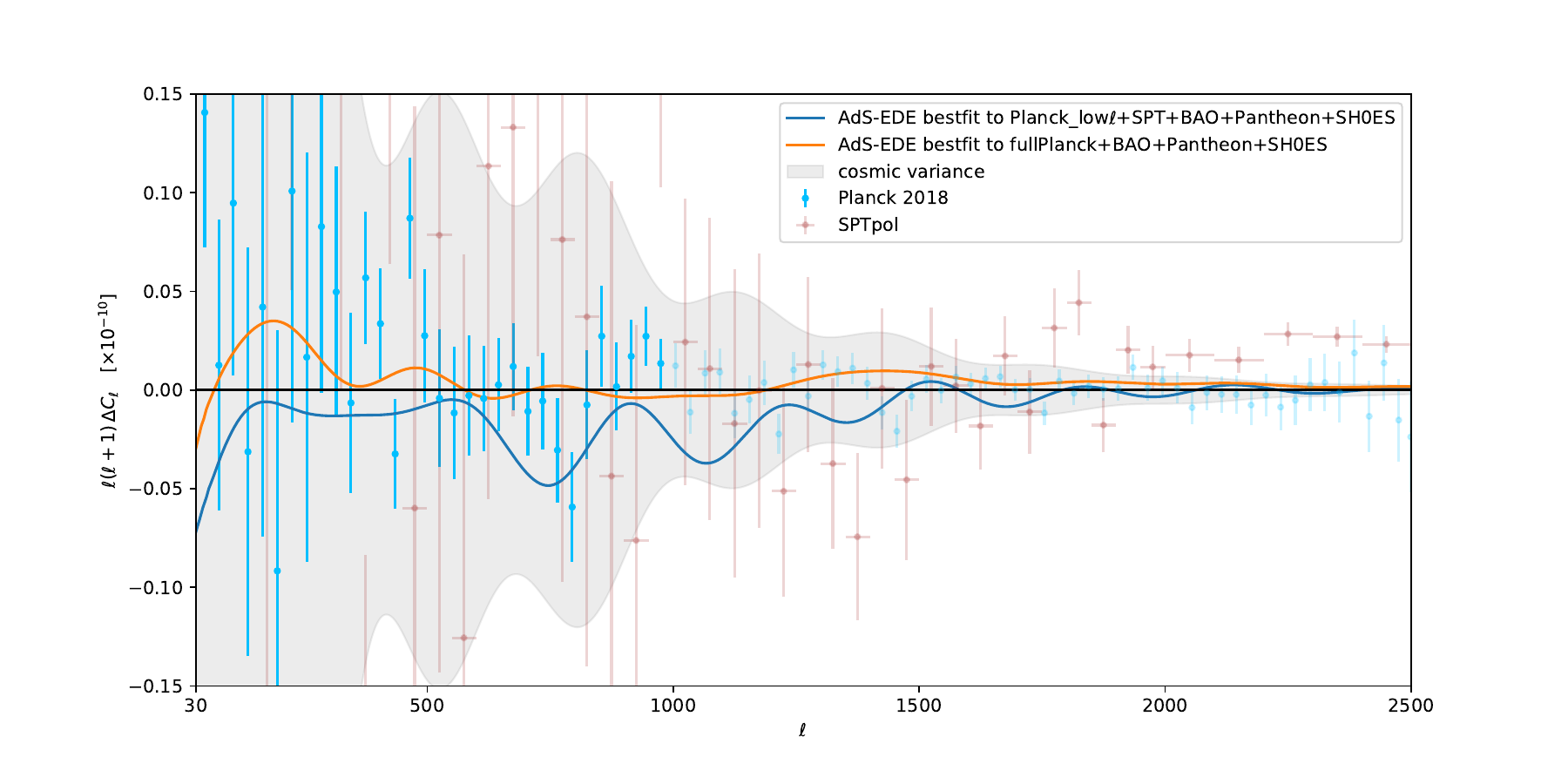}
\caption{\label{fig:AdS_TT}AdS-EDE best-fit values for
Planck\_low$\ell$+SPT+BAO+Pantheon+SH0ES and
fullPlanck+BAO+Pantheon+SH0ES on TT spectrum, and residual and
uncertainty of different CMB observations (light error bars is the
unused data). The reference value is the best-fit result to full
Planck data (i.e.
\texttt{base\_plikHM\_TTTEEE\_lowl\_lowE\_lensing}) under
$\Lambda$CDM model. Cosmological variance is showed in light gray
background. Note different $\ell$ range to other figures.}
\end{figure}

\begin{figure}[htb]
    \centering
   \includegraphics[width=\linewidth]{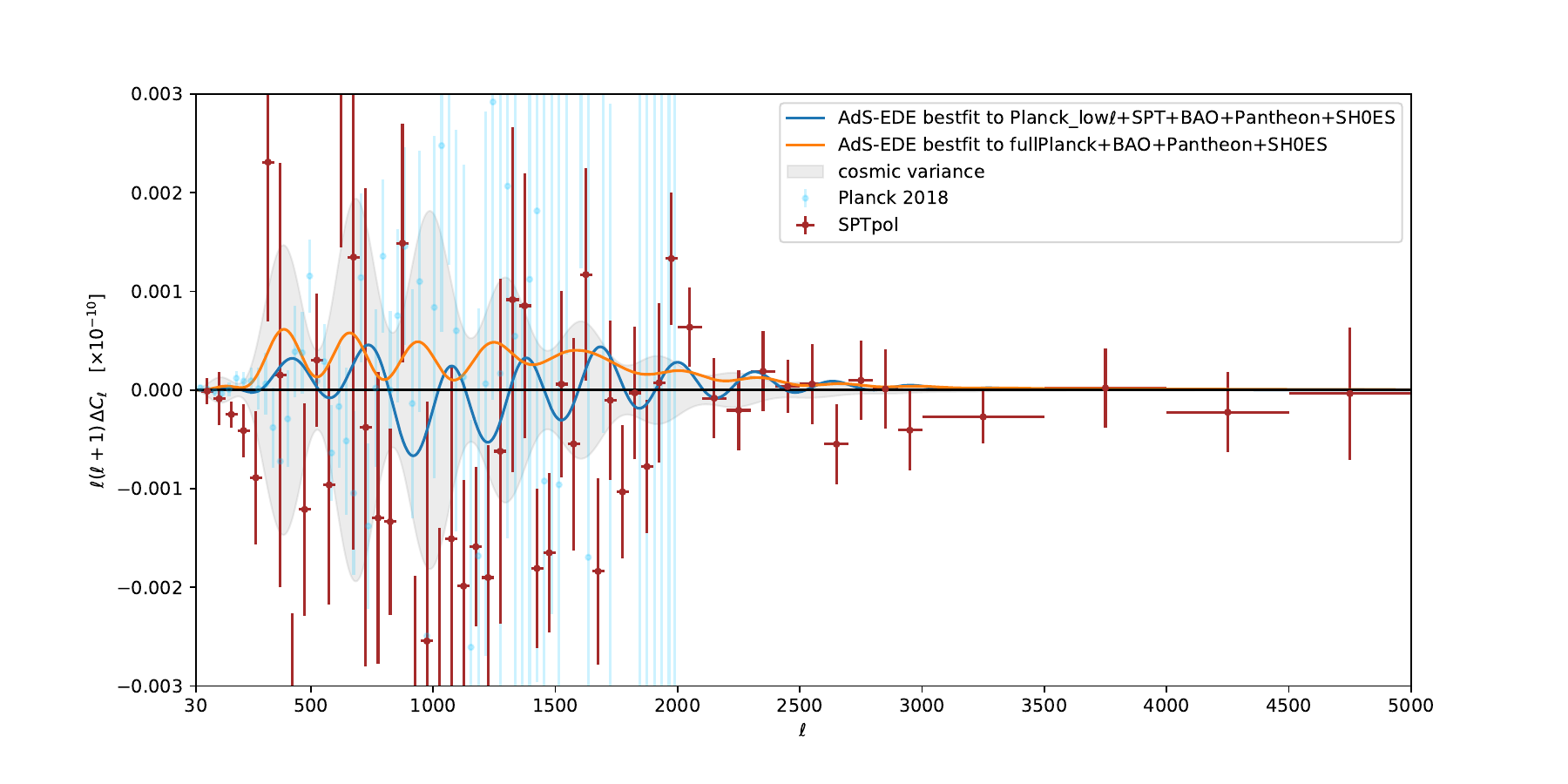}
\caption{\label{fig:AdS_EE}AdS-EDE best-fit values for
Planck\_low$\ell$+SPT+BAO+Pantheon+SH0ES and
fullPlanck+BAO+Pantheon+SH0ES on EE spectrum, and residual and
uncertainty of different CMB observations (light error bars is the
unused data). The reference value is the best-fit result to full
Planck data (i.e.
\texttt{base\_plikHM\_TTTEEE\_lowl\_lowE\_lensing}) under
$\Lambda$CDM model.}
\end{figure}

To clarify the effect of different CMB datasets, we plot their
best-fit values and residuals to Planck 2018 on TT and EE spectrum
in \autoref{fig:AdS_TT} and \autoref{fig:AdS_EE}. It can be found
in \autoref{fig:AdS_EE} that SPTpol polarization spectrum has
smaller uncertainty than Planck at $\ell>1000$ and can extend to
higher $\ell$. The different oscillation phases can be observed
and the fullPlanck case is closer to the baseline model (as our
baseline model contains fullPlanck data), which are indications
for different $\theta_s$. This different $\theta_s$ is due to the
oscillation-like residuals in Planck's TT spectrum, which cause
high-low $\ell$ inconsistency and lensing anomaly
\footnote{However, it is worth noting that our model also captures
$800 < \ell < 1000$ residual peaks of Planck TT.}. It also comes
from oscillation-like residuals in Planck's TE and EE spectrum,
which have not still detected in ACT or SPT observation
\cite{Lin:2020jcb}.


Regardless of which CMB dataset is used, compared with the
$\Lambda$CDM model larger $n_s$ brings a deficit in the $\ell<30$
multipoles. And this deficit does exist in Planck's observations
of the TT spectrum, which is one of the sources of its high and
low $\ell$ inconsistency \cite{Planck:2016tof}. This leads to a
better fit of AdS-EDE to this part of data, although not
significant due to extremely large cosmological variance here.

\begin{figure}[htb]
    \centering
   \includegraphics[width=\linewidth]{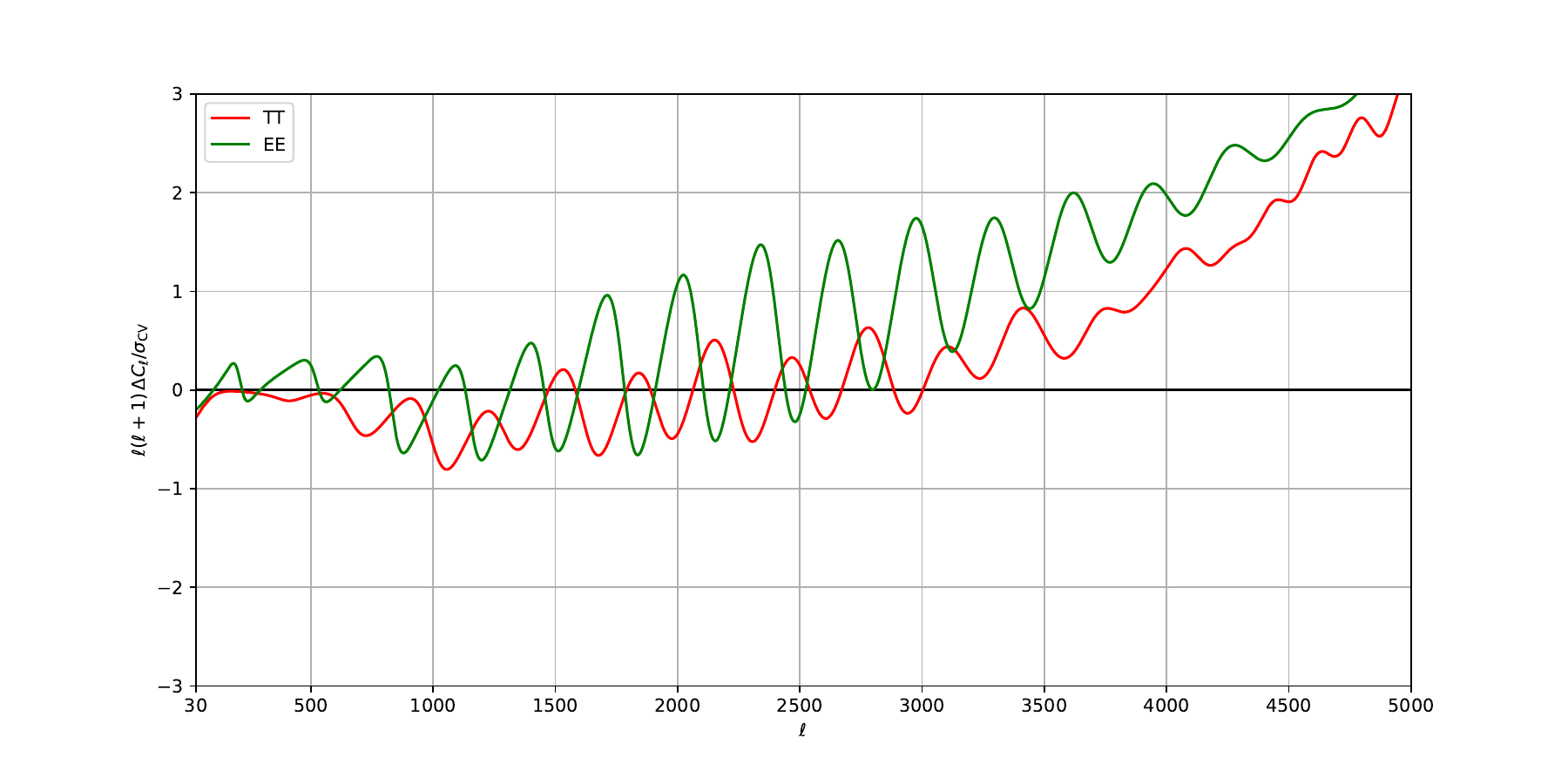}
\caption{\label{fig:AdS_TTEE_CV}The AdS-EDE best-fit values for
Planck\_low$\ell$+SPT+BAO+Pantheon+SH0ES compared with the
best-fit values for the $\Lambda$CDM model based only on full
Planck data (i.e.,
\texttt{base\_plikHM\_TTTEEE\_lowl\_lowE\_lensing}) in term of the
ratio of residuals to cosmological variance on the CMB TT and EE
spectra.}
\end{figure}

We also compare the residuals of AdS-EDE with the cosmological
variance, see \autoref{fig:AdS_TTEE_CV}. The residuals of the TT
spectrum can only be within the cosmological variance for scales
that can be observed at present ($\ell \leq 4000$), however, the
residuals of the EE spectrum is larger than the cosmological
variance at small scales ($2500 \lesssim \ell \leq 4000$). This
suggests that the polarization spectrum will be a significant tool
for constraining AdS-EDE from CMB, a range that also lies outside
Planck's multipoles. Due to the larger $n_s$, the AdS-EDE effect
on very small scales also far exceeds the cosmological variance.

\section{Conclusion}
\label{sec:7}

Currently, the EDE is a popular resolution for the Hubble tension.
In AdS-EDE model, the existence of AdS phase can make the energy
injected before recombination decay faster, which avoids degrading
the CMB fit, and thus allow a higher EDE fraction. However, like
other EDE models, this brings a larger late matter density
fluctuation, worsening the tension between Planck data and LSS
observations. Thus it is significant to revisit the constraining
power of LSS on the AdS-EDE model using the different CMB dataset.

It is well-known that the Planck data itself also present the
outstanding anomalies, such as the discrepancy between high and
low $\ell$ and the over-smoothing of the acoustic peak by the
gravitational lensing. Thus for CMB data, we conservatively
discard the high $\ell$ part of Planck TT spectrum as well as the
polarization spectrum and replace the corresponding data with
SPTpol observations, as in
Refs.\cite{Chudaykin:2020acu,Chudaykin:2020igl}. We verify the
compatibility of this combined CMB dataset under AdS-EDE.

We get $H_0 = 74.7$ km/s/Mpc when using our combined CMB dataset
alone (Planck\_low$\ell$+SPT), and $H_0 = 71.92$ km/s/Mpc when
using Planck\_low$\ell$+SPT+BAO+Pantheon dataset, respectively.
Here, the Hubble tension is substantially relieved, even without
any prior of direct measurements. \footnote{With the $H_0$ prior from SH0ES, we get $H_0=73.29$ km/s/Mpc. Inclusion of the RSD and
WL measurements leads to $S_8=0.785$, which is $1.3\sigma$
consistent with direct $S_8$ measurement.} And it is consistent with the
age constraints of the oldest astrophysical objects
\cite{Vagnozzi:2021tjv} (see also \cite{Jimenez:2019onw,Bernal:2021yli,Boylan-Kolchin:2021fvy,Krishnan:2021dyb} for other age constraints). Unlike other works (e.g.
\cite{Hill:2020osr,Chudaykin:2020igl}), we can see a preference
for the non-zero AdS-EDE, with the best-fit value
$\omega_\text{scf} \approx 0.1$. This is actually a straight
result of the AdS bound in AdS-EDE model, which is not present
without the AdS phase.

We also investigated the role of different CMB datasets. We find
that the direction of solving both $H_0$ tension and $S_8$ tension
is controlled by $\Omega_\text{m}$, which is related to
$\theta_s$. The oscillation-like residual at high-$\ell$ part of
Planck data, which is relative to lensing anomaly and high-low
$\ell$ inconsistency, results in the corresponding discrepancy of
$\theta_s$. The residual analysis shows that the small-scale
polarization spectrum is essential for identifying EDE models.
Ground-based CMB experiments such as SPT-3G \cite{ SPT-3G:2014dbx}
and Advanced ACTPol \cite{ Calabrese:2014gwa} might observe
effects of EDE on the CMB, while the upcoming high-precision
small-scale CMB experiments Simons Observatory
\cite{SimonsObservatory:2018koc} and CMB-S4 \cite {CMB-S4:2016ple}
can help us to reestimate the anomaly at Planck's high-$\ell$
part.

It should be also pointed out that the observations of LSS such as
Euclid satellite \cite{Amendola:2016saw}, LSST
\cite{LSSTDarkEnergyScience:2018jkl}, DESI \cite{DESI:2016fyo}
will not only further constraint on the resolution of the $H_0$
tension, but also enable us to judge whether the $S_8$ tension
actually implies new physics or not.

\acknowledgments

We thank Gen Ye for helpful discussions. This work is supported by
NSFC, Nos.12075246, 11690021, and also UCAS Undergraduate
Innovative Practice Project.

\appendix
\section{Results with Planck\_low$\ell$ + SPT + BAO + Pantheon + SH0ES + EFTofLSS}
\label{sec:A}

Effective Field Theory of Large-Scale Structure (EFTofLSS)
\cite{Baumann:2010tm,Carrasco:2012cv} can extract information on
moderate nonlinear scales, by characterizing the nonlinear effects
on small scales as effective parameters on large scales. Since LSS
is 3-dimensional, it will have larger number of modes than the
2-dimensional CMB, and thus can have a smaller cosmological
variance and theoretically better constrain cosmological
parameters.

Here, our BOSS DR12 BAO data are replaced with the data of BOSS
DR12 containing three regions \cite{Gil-Marin:2015sqa}: CMASS NGC,
CMASS SGC and LOWZ NGC, as well as the covariance matrix between
them. In principle, this statistical constraint on BAO data
selection is worse than the `consensus' final result used in our
text. We consider the similar model-independent approach to obtain
this BAO+Pantheon constraint on $r_s^\text{drag} H_0$:
$\beta_\text{BAO} = 29.684_{-0.386}^{+0.390}$. There is no
significant difference from the analysis result in the text, which
indicates that it is safe to use this BAO alternative. The
EFTofLSS is implemented by \texttt{pybird} \cite{DAmico:2020kxu}.
The nuisance parameter of EFTofLSS is partially marginalized, and
the rest are the same as \cite{DAmico:2020kxu}, independent on
echo area.

A comparison of the posterior distribution of the parameters
before and after including EFTofLSS is presented in
\autoref{fig:AdS_LSS} and \autoref{tab:+EFTofLSS}. It can be seen
that the degradation of the parameters is smaller than RSD+WL.
Including EFTofLSS, we have $H_0 = 72.34$ km/s/Mpc, the
discrepancy with R19 is only $0.8\sigma$, the $S_8$ is only
reduced by less than $1\sigma$, and the uncertainty does not
increase significantly.

\begin{table}[H]
\centering
\small
\centerline{\begin{tabular}{|c|c|c|}
\hline
parameters & Planck\_low$\ell$+SPT+BAO+Pantheon+SH0ES & Planck\_low$\ell$+SPT+BAO+Pantheon+SH0ES+EFTofLSS \\
\hline
 $\displaystyle \omega _{\text{scf}}$ & $\displaystyle 0.122( 0.116)_{-0.021}^{+0.014}$ & $\displaystyle 0.114( 0.108)_{-0.016}^{+0.010}$ \\
$\displaystyle \ln( 1+z_{c})$ & $\displaystyle 8.39( 8.30)_{-0.23}^{+0.17}$ & $\displaystyle 8.39( 8.50)_{-0.24}^{+0.19}$ \\
$\displaystyle H_{0}$ & $\displaystyle 73.08( 73.29) \pm 0.96$ & $\displaystyle 72.68( 72.34) \pm 0.91$ \\
$\displaystyle n_{s}$ & $\displaystyle 0.9965( 0.9990)_{-0.0065}^{+0.0083}$ & $\displaystyle 0.9938( 0.9945)_{-0.0063}^{+0.0079}$ \\
$\displaystyle \omega _{\text{b}}$ & $\displaystyle 0.02368( 0.02375)_{-0.00027}^{+0.00033}$ & $\displaystyle 0.02355( 0.02373)_{-0.00025}^{+0.00029}$ \\
$\displaystyle \omega _{\text{cdm}}$ & $\displaystyle 0.1316( 0.1308) \pm 0.0041$ & $\displaystyle 0.1304( 0.1292) \pm 0.0032$ \\
$\displaystyle \ln 10^{10} A_{s}$ & $\displaystyle 3.045( 3.046)_{-0.016}^{+0.021}$ & $\displaystyle 3.027( 3.030)_{-0.016}^{+0.028}$ \\
$\displaystyle \tau $ & $\displaystyle 0.0462( 0.0470)_{-0.0072}^{+0.011}$ & $\displaystyle 0.0382( 0.0434)_{-0.0077}^{+0.015}$ \\
\hline
 $\displaystyle 100\theta _{s}$ & $\displaystyle 1.03955( 1.03965) \pm 0.00073$ & $\displaystyle 1.03963( 1.04010) \pm 0.00070$ \\
$\displaystyle \sigma _{8}$ & $\displaystyle 0.831( 0.832)_{-0.013}^{+0.017}$ & $\displaystyle 0.821( 0.819)_{-0.013}^{+0.015}$ \\
$\displaystyle S_{8}$ & $\displaystyle 0.820( 0.816)_{-0.017}^{+0.020}$ & $\displaystyle 0.811( 0.810) \pm 0.016$ \\
 \hline
\end{tabular}}
\caption{\label{tab:+EFTofLSS}Constraints on the parameters of
AdS-EDE for Planck\_low$\ell$ + SPT + BAO + Pantheon + SH0ES (+
EFTofLSS), including mean values and $\pm 1\sigma$ regions, with
best-fit values in parentheses.}
\end{table}

\section{Results with fullPlanck+BAO+Pantheon}
\label{sec:B}

\begin{table}[H]
\centering
\centerline{\begin{tabular}{|c|c|c|}
\hline
parameters & fullPlanck+BAO+Pantheon\\
\hline
 $\displaystyle \omega _{\text{scf}}$ & $0.1124(0.1084)^{+0.0038}_{-0.0070}$ \\
$\displaystyle \ln( 1+z_{c})$ & $8.153(8.147)^{+0.075}_{-0.084}$ \\
$\displaystyle H_{0}$ & $72.52(72.46)\pm 0.51$ \\
$\displaystyle n_{s}$ & $0.9964(0.9949)^{+0.0047}_{-0.0041}$ \\
$\displaystyle \omega _{\text{b}}$ & $0.02341(0.02331)^{+0.00018}_{-0.00016}$ \\
$\displaystyle \omega _{\text{cdm}}$ & $0.1346(0.1336)^{+0.0016}_{-0.0018}$ \\
$\displaystyle \ln 10^{10} A_{s}$ & $3.079(3.072)\pm 0.015$ \\
$\displaystyle \tau $ & $0.0545(0.0523)^{+0.0071}_{-0.0079}$ \\
\hline
 $\displaystyle 100\theta _{s}$ & $1.04108(1.04132)\pm 0.00029$ \\
$\displaystyle \sigma _{8}$ & $0.8604(0.8554)\pm 0.0074$ \\
$\displaystyle S_{8}$ & $0.863(0.856)\pm 0.011$ \\
 \hline
\end{tabular}}
\caption{\label{tab:fullPlanck}Constraints on the parameters of
AdS-EDE for fullPlanck+BAO+Pantheon, including mean values and
$\pm 1\sigma$ regions, with best-fit values in parentheses.}
\end{table}

\bibliographystyle{JHEP}
\bibliography{./refs.bib}
\end{document}